\newif\ifdocutohuawei
\docutohuaweifalse

\ifdocutohuawei
\documentclass[conference]{IEEEtran}
\else
\documentclass[conference]{IEEEtran}
\fi
\usepackage{amsmath}
\usepackage{tabularx}  % 提供 tabularx 环境
\usepackage{url}       % 处理 URL 换行
\usepackage{booktabs}  % 提供更好的表格线（可选）
\usepackage{multirow}
\usepackage{graphicx}
\usepackage{array}

\ifCLASSINFOpdf
\else
\fi

\newcolumntype{C}[1]{>{\centering\arraybackslash}p{#1}}

\begin{document}

\ifdocutohuawei
\title{Manual V1.0: Gathering 802.11ax Channel State Information from ZTE Access Points}
\else
\title{Wi-Fi Sensing Tool Release: Gathering 802.11ax Channel State Information from a Commercial Wi-Fi Access Point}
\fi

\ifdocutohuawei
\author{Zisheng Wang
\thanks{ZTE Corporation，Shenzhen, China, e-mail: (wang.zisheng@zte.com.cn).}
}
\else
\author{Zisheng Wang$^{1}$, Feng Li$^{2}$, Hangbin Zhao$^{2}$, Zihuan Mao$^{1}$, Yaodong Zhang$^{1}$, Qisheng Huang$^{1}$, \\ Bo Cao$^{1}$, Mingming Cao$^{1}$, Baolin He$^{1}$, Qilin Hou$^{1}$
\thanks{$^{1}$Z. Wang, Y. Zhang, Q. Huang, B. Cao, M. Cao, B. He and Q. Hou, are with State Key Laboratory of Mobile Network and Mobile Multimedia Technology ，Shenzhen, China, e-mail: (wang.zisheng@zte.com.cn). Authors are ordered according to the contribution to this project.
}
\thanks{$^{2}$F. Li and H. Zhao are with China Mobile (Hangzhou) Information Technology (Co., Ltd)}
}
\fi

% \markboth{Journal of \LaTeX\ Class Files,~Vol.~14, No.~8, August~2015}%
% {Shell \MakeLowercase{\textit{et al.}}: Bare Demo of IEEEtran.cls for IEEE Journals}



\maketitle


\ifdocutohuawei
\else
\begin{abstract}
Wi-Fi sensing has emerged as a powerful technology, leveraging channel state information (CSI) extracted from wireless data packets to enable diverse applications, ranging from human presence detection to gesture recognition and health monitoring. 
However, CSI extraction from commercial Wi-Fi access point lacks and out of date. 
This \ifdocutohuawei
manual
\else
paper
\fi introduces ZTECSITool, a toolkit designed to capture high-resolution CSI measurements from commercial Wi-Fi 6 (802.11ax) access points, supporting bandwidths up to 160 MHz and 512 subcarriers. 
ZTECSITool bridges a critical gap in Wi-Fi sensing research, facilitating the development of next-generation sensing systems.
The toolkit includes customized firmware and open-source software tools for configuring, collecting, and parsing CSI data, offering researchers a robust platform for advanced sensing applications. 
We detail the command protocols for CSI extraction, including band selection, STA filtering, and report configuration, and provide insights into the data structure of the reported CSI. Additionally, we present a Python-based graphical interface for real-time CSI visualization and analysis. 
\end{abstract}
\fi

\ifdocutohuawei
\else
\fi

\IEEEpeerreviewmaketitle

\section{Introduction}

\ifdocutohuawei
\else
\IEEEPARstart{W}{i-fi} Sensing has been rapidly studied, using channel state information (CSI) obtained from the wireless data packets to enable a variety of applications from the most basic person present detection to gesture recognition and human heath monitoring. 
\fi 
We proposed a toolkit (ZTECSITool) that records CSI measurements with a recent 802.11ax protocol and a larger number of subcarriers (up to 512 for 160MHz). The toolkit use the ZTE AX3000 series Wi-Fi Access Point (AP) products \footnote{Two types of AX3000 AP are supported. Production Type E2631 for AX3000 and ZXSLC SR6110 for AX3000 Pro. Both APs have the same Wi-Fi parameters where AX3000 Pro (SR6110) has two 2.5G ETH ports.} with 3x Tx/Rx antennas for 5GHz and 2x Tx/Rx antenna for 2.4GHz. We provide a CSI toolkit to setup, record, and parse the CSI measurements for later analysis. 
\ifdocutohuawei
In this manual,
\else
In this paper, 
\fi
we also provide details of controlling methods and CSI structures so that users can design their own CSI applications.

\ifdocutohuawei
\else
\section{CSI Extraction Tool and Datasets}

In current CSI-based Wi-Fi sensing technique, researchers typically utilize open-source CSI datasets and toolkits from various commercial Wi-Fi network adapters. In this section, we briefly introduce the widely used Wi-Fi CSI extraction tools and open-source CSI datasets for Wi-Fi sensing.

\subsection{CSI Extraction Tool}

We first compiled a list of commonly used CSI extraction tools in current research, which is presented in Table \ref{tbl:csitoolcompare}. The currently available and most widely used open-source CSI extraction tools includes  CSITool \cite{Halperin2011Tool}, Nexmon CSI Extractor \cite{Schulz2021Nexmon}, Atheros CSI Tool \cite{Xie2015Precise}, ESP32 CSI Toolkits \cite{Hernandez2020Lightweight}, and the proposed ZTECSITool introduced in this work. A detailed comparison of the above tools, including Wi-Fi chipset architectures, operating system platforms, multi-input multi-output (MIMO) characteristics, 802.11 support, and additional parameters, is provided in the table. It is noteworthy that, compared to other CSI extraction tools, our proposed MT7916-based ZTECSITool supports both Linux and Windows Operating System platform and is compatible with the IEEE 802.11ax (WiFi 6) standard. Based on the 802.11ax standard and the advanced chipset design, ZTECSITool supports a bandwidth of up to $160\;\mathrm{MHz}$ and $512$ subcarriers quantized with 16 bits. This capability provides more extensive and precise CSI information for Wi-Fi sensing.

\begin{table}[!t]
\caption{Current CSI extraction tools compared with ZTECSITool}
\label{tbl:csitoolcompare}
\centering
\renewcommand{\arraystretch}{1.5} % 增大行间距
\begin{tabular}{p{1.5cm} p{0.7cm} p{1cm} p{0.7cm} p{1.2cm} p{1cm}}
\hline\hline
Name  & MIMO & 802.11 Support & Sub-carriers & Bandwidth (MHz) & Resolution (bits) \\
\hline
\raggedright CSITool & 2x2 & 11n & 60 & 40 & 8 \\
\raggedright Nexmon CSI Extractor &  4x4 & 11n/ac & & 80 & \\
\raggedright Atheros CSI Tool & - & 11n & 114 & 40 & 8 \\
ZTECSITool &  3x2 & 11n/ac/ax & 512 & 160 & 16 \\
\hline\hline
\end{tabular}
\end{table}

% Table list 
\subsection{Opensource CSI Dataset}
As WiFi sensing research advances, numerous open-source CSI datasets have become available. We categorize these datasets into three main types, based on their collection methods and data scale.

The first type of CSI dataset is collected by Wi-Fi systems with distributed antennas, e.g., WiSDAR \cite{Wang2019On}, WiNDR \cite{Qin2024Direction} and WiCross \cite{Qin2023Cross}. Concretely, WiSDAR evaluates different distributed antenna topologies, such as line, hexagon, square, and random arrangements, which reveals the regions most sensitive to human activities and achieves both high accuracy and reliability in Wi-Fi-based recognition. In order to design the placement of distributed antennas based on the principles of electromagnetic propagation, WiNDR and WiCross place three antennas for both the transmitter and receiver in a staggered and distributed manner around a $\mathrm{360}^{\circ}$ circle. Such configuration can effectively alleviate the confusion caused by the orientation according to the characteristics of the Fresnel region, and thus, enable the direction-agnostic gesture recognition. 

% WiNDR \cite{Qin2024Direction} and WiCross \cite{Qin2023Cross} position three antennas from the transmitter and three antennas from the receiver in a staggered, distributed manner around a 360-degree circle, with the subject performing gestures at the center. This configuration achieves direction-agnostic gesture recognition. Additionally, WiSDAR [194] tests various distributed antenna topologies, including line, hexagon, square, and random shapes, to explore regions that are most sensitive to human activities, achieving highly accurate and reliable recognition results.

The second type utilizes additional Wi-Fi devices to create a distributed system for CSI extraction. For example, the scenarios of single transmitter with multiple receivers are investigated in ReWiS \cite{Bahadori2022ReWiS}, WiTraj \cite{Wu2021WiTraj}, Widar 3.0 \cite{Zheng2019Zero} and OneFi \cite{Xiao2021One}. Through this distributed observation approach, the essential environment-independent information required for Wi-Fi sensing can be collected via CSI. Additionally, the theoretical relationship between sensing range and the transceivers distance is derived by Wang et al \cite{Wang2022Placement}, which reveals that optimizing the layout of distributed devices can effectively enhance the sensing coverage and suppress environmental interference on sensing.
The last type is the dataset with scaling up training data. In the context of data-driven Wi-Fi sensing methods, increasing the amount of training data is a vital method for enhancing the generalization capabilities of deep learning approaches. To this end, The representative methods, including Widar3.0 \cite{Zheng2019Zero}, MM-Fi \cite{Yang2022Advances}, XRF55 \cite{Wang2024XRF55} and CSI-PCNH \cite{Zhang2023CSI}, have all acquired large-scale CSI data for training. In the above work, the researcher researchers have found that training on tens of thousands of samples allows the sensing model to naturally achieve invariance to direction and position, while also supporting adaptation to new environments through learning from a small number of samples and fine-tuning model parameters.

\fi

% Intel CSI Tools

% Nexmon: The C-based Firmware Patching Framework
% Channel State Information extractor for various Wi-Fi chips
% It allows to extract CSI of up to 4x4 MIMO transmissions at 80 MHz bandwidth

% Atheros CSI Tool
% https://wands.sg/research/wifi/AtherosCSI/

% ESP32 CSI Toolkit
% https://stevenmhernandez.github.io/ESP32-CSI-Tool/

\section{Details on ZTECSITool}

In this section, we describe 
\ifdocutohuawei
\else
our contribution, 
\fi
ZTECSITool, which allows WLAN sensing researchers collect CSI raw data up to 512 sub-carriers for each chain from our commercial ZTE AX3000 APs. ZTECSITool consists of
\begin{itemize}
    \item ZTE AX3000 Series AP: which accepts CSI configuration commands and uploads CSI raw data from and to LAN/WLAN.
    \item A PC software: which helps to control and parse the CSI raw data.
\end{itemize}
We first present the system architecture. Then, the CSI control commands and CSI data format are described in detail. Next, we present ZTECSITool software. Finally, essential usage notes for ZTECSITool are outlined. 
Anyone can access the latest information and access ZTECSITool from https://github.com/WiFiZTE2025/ZTE\_WiFi\_Sensing.git. 

\subsection{System Architecture}

ZTECSITool maximizes usability by taking into account the diverse needs of different research scenarios. Hence, a CSI collection system using ZTECSITool consist of four parts:
\begin{itemize}
    \item ZTE AP: The key part of the system. It estimates CSI from received PPDU preambles and uploads CSI accordingly. 
    \item A Wi-Fi STA: usually a mobile phone. In WLAN sensing applications, Wi-Fi STAs typically transmit PPDUs based on specific constraints or protocol-defined requirements.
    \item A CSI Controller: usually a computer. It sends CSI commands to the ZTE AP. It equips the software in ZTECSITool. 
    \item A CSI Collector: usually a computer. It receives CSI from the ZTE AP after the AP is configured by the CSI controller. 
\end{itemize}
\begin{figure}[htbp]
    \centering
    \includegraphics[width=0.45\textwidth]{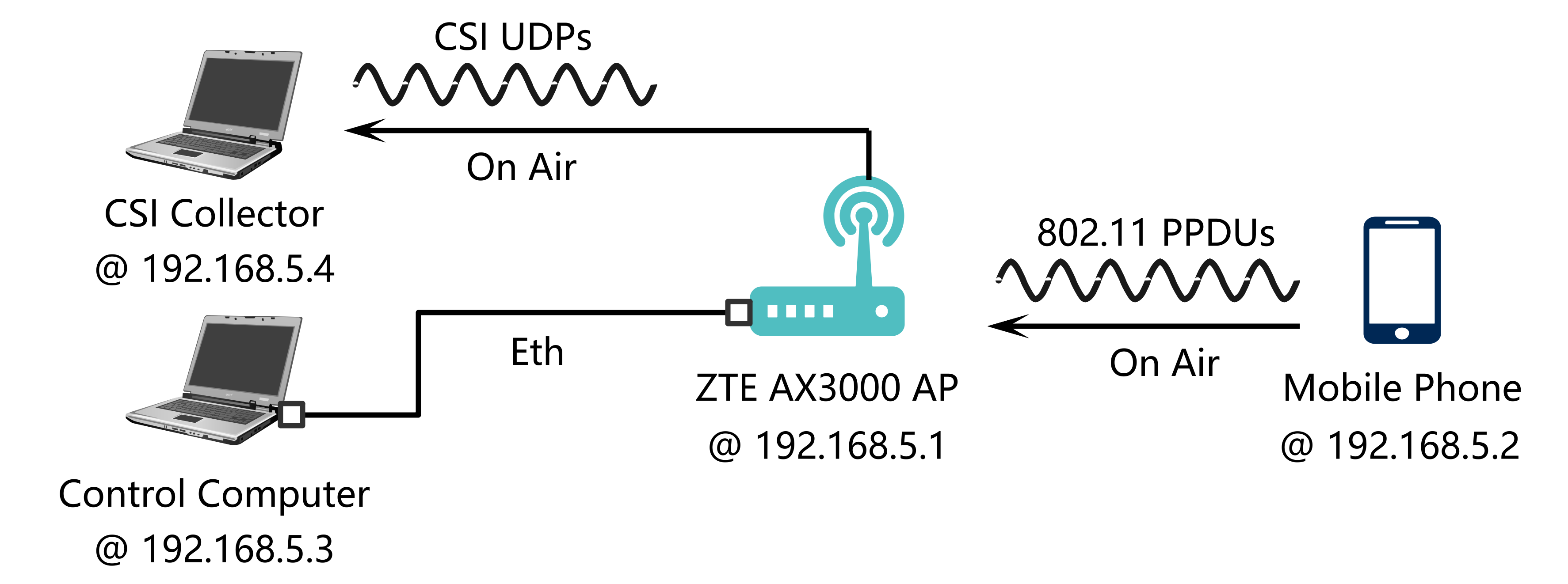}
    \caption{An example system architecture of the CSI collection system using ZTECSITool.}
    \label{fig:sys}
\end{figure}
Figure \ref{fig:sys} shows an example system architecture of the CSI collection system using ZTECSITool.
For the convenience, CSI Controller and Collector talks with the ZTE AP on IP protocol so that CSI controller and collector can use LAN or WLAN to finish their job once it obtains IP addresses from ZTE AP. The Controller and Collector can be the same computer. Note that, controller and collector require IP addresses assigned by ZTE AP, so it can't connect with ZTE AP using WAN. Which LAN port is WAN is subject to the user configuration. 
In the next section, we introduce how the CSI Controller should operate and how to issue CSI control commands.

\subsection{CSI Commands}\label{sec:config}
CSI Controller sends CSI command by using pre-defined UDP packets to the gateway of ZTE AP with port 8021.
The UDP packets contains a magic number, command type and command data which is listed in TABLE \ref{tbl:udpcmdformat}.
ZTE AP listens UDP packets at 8021. Once ZTE AP parses correct magic number from captured UDP packets, it will start to parse the whole command and execute the corresponding job as per defined in this section. 
All field is little-end. The definition of CMD\_DATA field depends on CMD\_TYPE.

\begin{table}[!t]
\caption{The overall format of the CSI configuration UDP command}
\label{tbl:udpcmdformat}
\centering
\renewcommand{\arraystretch}{1.5} % 增大行间距
% Some packages, such as MDW tools, offer better commands for making tables
% than the plain LaTeX2e tabular which is used here.
\begin{tabular}{p{1.5cm}p{0.5cm}p{1cm}p{4cm}}
\hline\hline
field      & count   & type      & description                                                                           \\
\hline
MAGIC\_NUM & 1       & uint64\_t & The magic num is set to 0xCAFE2025. ZTE AP use this field to validate the UDP packets \\
CMD\_TYPE  & 1       & uint8\_t  & The type of cmd.                                                                      \\
CMD\_DATA  & various & -         & CMD\_DATA depends on the type of cmd.                                                 \\
\hline\hline
\end{tabular}
\end{table}

\subsubsection{CSI Report Enable Command (0x1)}
By setting CMD\_TYPE to 0x1 in the CSI Configuration UDP command, we can enable or disable the CSI report. The format of CSI Report Enable Command is defined in TABLE \ref{tbl:csirpten}.

\begin{table}[!t]
\caption{The format of the CSI Report Enable Command}
\label{tbl:csirpten}
\centering
\renewcommand{\arraystretch}{1.5} % 增大行间距
\begin{tabular}{p{1.5cm}p{0.5cm}p{1cm}p{4cm}}
\hline\hline
field      & count & type      & description                                                                           \\
\hline
MAGIC\_NUM & 1     & uint64\_t & The magic num is set to 0xCAFE2025. ZTE AP use this field to validate the UDP packets \\
CMD\_TYPE  & 1     & uint8\_t  & Set to 0x1 for CSI Report Enable Command                                              \\
Enable     & 1     & uint8\_t  & 0 - for stopping CSI report                                                           \\
           &       &           & 1 - for starting CSI report \\                                     
\hline\hline
\end{tabular}
\end{table}

\subsubsection{CSI STA Filter Command (0x2)}
ZTE AP might have many STA associated. By default, if CSI reporting is enabled, ZTE AP will report CSI based on PPDUs from every STA. Using CSI STA Filter Command, ZTE AP will only report CSI data from certain STAs. ZTE AP supports at most 5 STAs in the filter list. ZTE AP will not check if filtered STA has associated or if the STA mac is valid. The format of CSI STA Filter Command is defined in TABLE \ref{tbl:csistafilter}

\begin{table}[!t]
\caption{The format of the CSI STA Filter Command}
\label{tbl:csistafilter}
\centering
\renewcommand{\arraystretch}{1.5} % 增大行间距
\begin{tabular}{p{1.5cm}p{0.5cm}p{1cm}p{4cm}}
\hline\hline
field      & count & type      & description \\\hline
MAGIC\_NUM & 1     & uint64\_t & The magic num is set to 0xCAFE2025. ZTE AP use this field to validate the UDP packets \\
CMD\_TYPE  & 1     & uint8\_t  & Set to 0x2 for CSI STA Filter Command. Suppose that the MAC address is 01:02:03:04:05:06 \\
STA MAC 0  & 1     & uint8\_t  & then, STA MAC 0 = 0x01 \\
STA MAC 1  & 1     & uint8\_t  & then, STA MAC 1 = 0x02 \\
STA MAC 2  & 1     & uint8\_t  & then, STA MAC 2 = 0x03 \\
STA MAC 3  & 1     & uint8\_t  & then, STA MAC 3 = 0x04 \\
STA MAC 4  & 1     & uint8\_t  & then, STA MAC 4 = 0x05 \\
STA MAC 5  & 1     & uint8\_t  & then, STA MAC 5 = 0x06 \\                                            
\hline\hline     
\end{tabular}
\end{table}

\subsubsection{CSI Configuration Command (0x3)}
ZTE AP can report CSI of PPDUs form certain types. We can also configure the number of chains reported. The format of CSI Configuration Command is defined in TABLE \ref{tbl:csiconfg}

\begin{table}[!t]
\caption{The format of the CSI Configuration Command}
\label{tbl:csiconfg}
\centering
\renewcommand{\arraystretch}{1.5} % 增大行间距
\begin{tabular}{p{1.5cm}p{0.5cm}p{1cm}p{4cm}}
\hline\hline
field            & count & type      & description \\\hline
MAGIC\_NUM       & 1     & uint64\_t & The magic num is set to 0xCAFE2025. ZTE AP use this field to validate the UDP packets  \\
CMD\_TYPE        & 1     & uint8\_t  & Set to 0x3 for CSI Configuration Command.  \\
FRAME\_TYPE & 1     & uint8\_t  & The FRAME\_TYPE are defined as per 9.2.4.13 Type and Subtype subfields (Draft P802.11 REVme\_D7.0). For example, as per Table 9-1, QoS data is indicated by \\
        & & & B3 B2 = 2b'10 \\
        & & & B7 B6 B5 B4 = 4b'1000\\
        & & & Then FRAME\_TYPE shall be set to \\
        & & & 0 0 B7 B6 B5 B4 B3 B2 = 0x22\\
        & & & Note that ZTE won't guarantee that ppdus with any frame type can report valid CSI. We encourage user use QoS data (FRAME\_TYPE = 0x22)\\
\hline\hline     
\end{tabular}
\end{table}

\subsubsection{CSI Report Configuration Command (0x4)} 
By default, CSI is reported by fragmented UDP packets sending to destination IP 192.168.X.2, destination port 8023 and source port 8024.The X in the destination IP depends on how the gateway of ZTE AP is set. But we allow user to change the destination IP to whichever they want. If this IP address is assigned to a device which is a 802.11 STA connects to the ZTE AP, the CSI will be reported through WLAN inferface. If this IP address is assigned to a devices which, for example, is a PC which connects to the ZTE AP by LAN, the CSI will be reported through the LAN connected. The format of CSI Report Configuration Command is defined in TABLE \ref{tbl:csireport}.
 
\begin{table}[!t]
\caption{The format of the CSI Report Configuration Command}
\label{tbl:csireport}
\centering
\renewcommand{\arraystretch}{1.5} % 增大行间距
\begin{tabular}{p{1.5cm}p{0.5cm}p{1cm}p{4cm}}
\hline\hline
field      & count & type      & description                                                                           \\\hline
MAGIC\_NUM & 1     & uint64\_t & The magic num is set to 0xCAFE2025. ZTE AP use this field to validate the UDP packets \\
CMD\_TYPE  & 1     & uint8\_t  & Set to 0x4 for CSI Report Configuration Command.                                             \\
TGT\_IP    & 4     & uint8\_t  & if target ip address is 192.168.1.1 then   \\
& & & TGT\_IP[0] = 192\\
& & & TGT\_IP[1] = 168\\
& & & TGT\_IP[2] = 1\\
& & & TGT\_IP[3] = 1\\
\hline\hline     
\end{tabular}
\end{table}

\subsubsection{CSI Band Configuration Command (0x5)}

The user can determine whether CSI is collected from 2.4G band or 5G band. Due to the limitation of system design, ZTE AP only supports one band a time. The format of CSI Band Command is defined in TABLE \ref{tbl:csiintf}.

\begin{table}[!t]
\caption{The format of the CSI Band Configuration Command}
\label{tbl:csiintf}
\centering
\renewcommand{\arraystretch}{1.5} % 增大行间距
\begin{tabular}{p{1.5cm}p{0.5cm}p{1cm}p{4cm}}
\hline\hline
field      & count & type      & description                                                                           \\\hline
MAGIC\_NUM & 1     & uint64\_t & The magic num is set to 0xCAFE2025. ZTE AP use this field to validate the UDP packets \\
CMD\_TYPE  & 1     & uint8\_t  & Set to 0x5 for CSI Band Configuration Command                                    \\
BAND  & 1     & uint8\_t  & 0 for 2.4G band  \\
& & & 1 for 5G band\\
\hline\hline     
\end{tabular}
\end{table}

\subsubsection{Check Availability Command (0x6)}

We allow user to check if ZTE AP is ready for CSI report. After received Check Availability Command, ZTE AP will reply an UDP packet with "OK" in the data field if ZTE AP is ready for CSI Report. The format of Check Availability Command is defined in TABLE \ref{tbl:chkcmd}

\begin{table}[!t]
\caption{The format of the Check Availability Command}
\label{tbl:chkcmd}
\centering
\renewcommand{\arraystretch}{1.5} % 增大行间距
\begin{tabular}{p{1.5cm}p{0.5cm}p{1cm}p{4cm}}
\hline\hline
field      & count & type      & description                                                                           \\\hline
MAGIC\_NUM & 1     & uint64\_t & The magic num is set to 0xCAFE2025. ZTE AP use this field to validate the UDP packets \\
CMD\_TYPE  & 1     & uint8\_t  & Set to 0x6 for Check Availability Command                                    \\
\hline\hline     
\end{tabular}
\end{table}

\subsection{How to use command}\label{sec:howto}

In Section \ref{sec:config}, we introduce the commands supported by ZTE AP to start CSI report. These commands should be sent to ZTE AP in a required order so that ZTE AP can work correctly. To start the CSI report, the required order is
\begin{enumerate}
    \item CSI Band Configuration Command (0x5)
    \item CSI Configuration Command (0x3)
    \item CSI Report Enable Command (0x1)
    \item CSI STA Filter Command (0x2)
    \item CSI Report Configuration Command (0x4)
\end{enumerate}
At least 500ms interval is required between each command. If there is STA sending PPDUs to ZTE AP, after step 3, we will observe fragmented UDP packets from ZTE AP. A few rules should be obeyed,
\begin{itemize}
    \item Steps 1–3 must be performed consecutively and not executed separately, as doing so may lead to unexpected errors.
    \item Only one band collection is supported at one time. If we already collect CSI from one band, before switching to another band, system reboot is required. 
\end{itemize}

\subsection{CSI Data Format}

Once the ZTE AP is setup properly as per Section \ref{sec:howto}, the raw CSI will be reported by UDP packets. Depends on the maximum transmission unit (MTU), the UDP packets might be fragmented. The users can listen on port 8023 with their local ip address. The CSI is saved in the data field of the UDP packets with little-end type. The format of CSI is defined in TABLE \ref{tbl:csidataformat}.

\begin{table}[!t]
\caption{The format of the CSI data}
\label{tbl:csidataformat}
\centering
\renewcommand{\arraystretch}{1.5} % 增大行间距
\begin{tabular}{p{1.5cm}p{0.5cm}p{1cm}p{4cm}}
\hline\hline
field      & count & type      & description                                                     \\\hline
MAGIC\_NUM & 1     & uint32\_t & The high 2 bytes is 0xCAFE.                                     \\
vendor     & 1     & uint8\_t  & For ZTE AX3000, this value is 2                                 \\
chip\_id   & 1     & uint32\_t & For ZTE AX3000, this value is 1                                 \\
timestamp & 1     & uint64\_t & Timestamp when the CSI is collected in us.                      \\
resv       & 1     & uint32\_t & resvered for later use                                          \\
bw         & 1     & uint32\_t & The bandwidth of the PPDUs. 0-4 represent 20MHz to 160MHz       \\
phy\_mode  & 1     & uint32\_t &                                                                 \\
resv\_1    & 1     & uint8\_t  & resevered for later use                                         \\
resv\_2    & 1     & uint16\_t & resevered for later use                                         \\
peer\_addr & 6     & uint8\_t  & The mac address of the PPDU which the CSI is estimated based on \\
rssi       & 16    & int32\_t  & The rssi of the PPDU                                            \\
resv\_3    & 16    & int32\_t  & reserved for later use                                          \\
agc\_gain  & 16    & int8\_t   & The RF AGC gain when the PPDU is received.                      \\
mcs        & 1     & int16\_t  & The MCS of the received PPDU                                    \\
gi\_type   & 1     & int8\_t   & The Guard Interval of the received PPDU                         \\
coding     & 1     & int8\_t   & The coding type of the received PPDU                            \\
stbc       & 1     & int8\_t   & Whether the received PPDU is coded by stbc                      \\
resv\_4    & 1     & int8\_t   & reserved for later use                                          \\
dcm        & 1     & int8\_t   & Whether the received PPDU uses dual-carrier modulation          \\
resv\_5    & 1     & int8\_t   & reserved for later use                                          \\
resv\_6    & 1     & uint64\_t & reserved for later use \\
csi\_cnt   & 1     & int16\_t  & The number of CSI in this packets                               \\
csi\_i     & 512   & int32\_t  & The I part of CSI  \\
csi\_q     & 512   & int32\_t  & The Q part of CSI  \\
\hline\hline     
\end{tabular}
\end{table}

\subsection{Some Examples on Commands and CSI data}

\begin{figure}[htbp]
    \centering
    \includegraphics[width=0.45\textwidth]{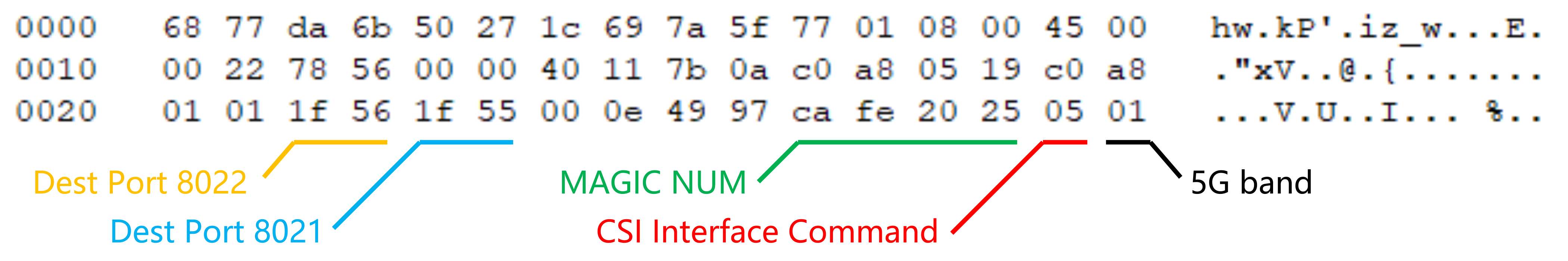}
    \caption{An example of an UDP packet carrying CSI Band Command}
    \label{fig:csiintf}
\end{figure}

In Figure \ref{fig:csiintf}, we provide an example of a UDP packet which carries a CSI Band Command captured by Wireshark. The CSI Band Command requires ZTE AP report CSI from 5G band.

\begin{figure}[htbp]
    \centering
    \includegraphics[width=0.45\textwidth]{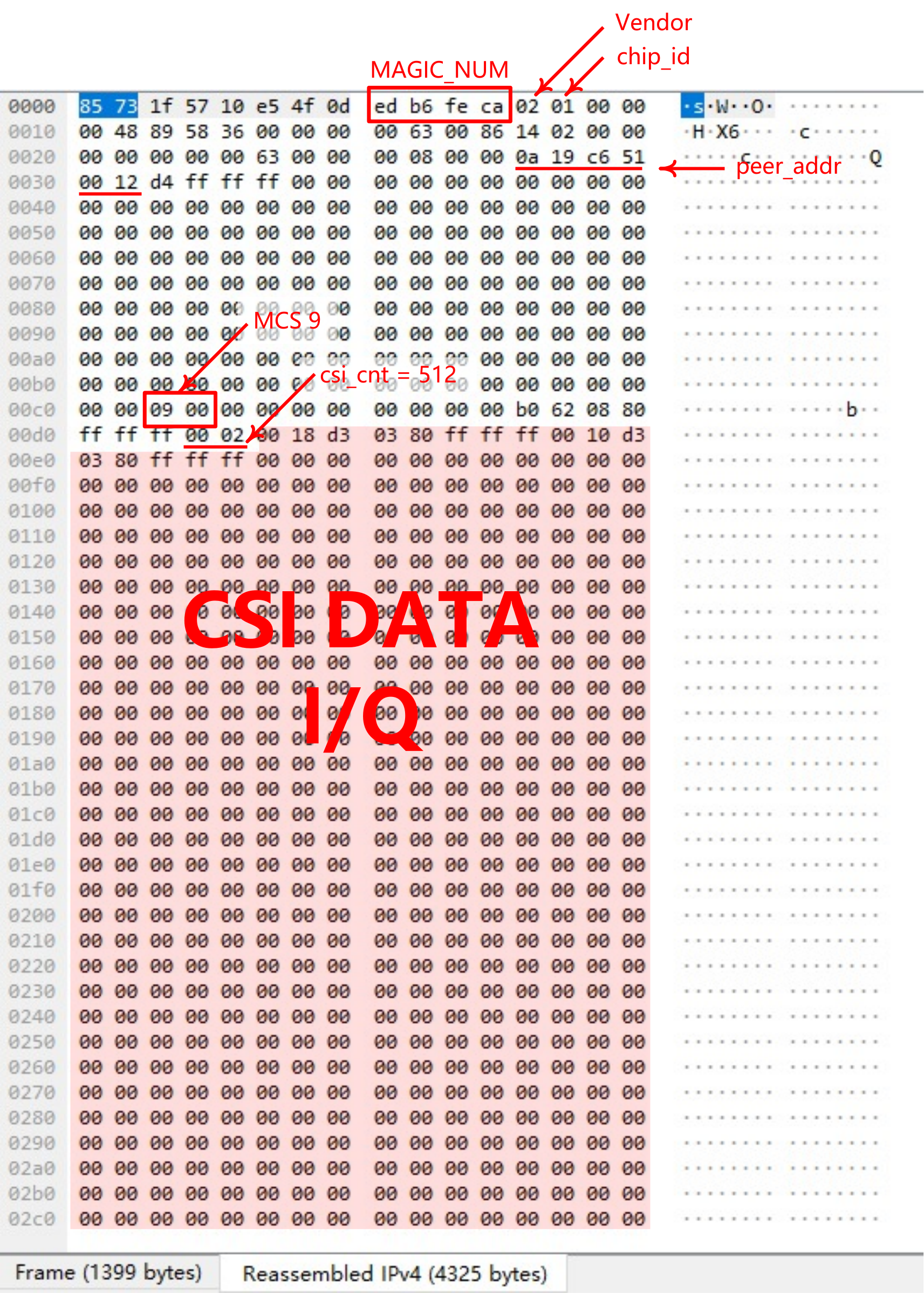}
    \caption{An example of an UDP packet carrying CSI data format.}
    \label{fig:csidataformat}
\end{figure}

In figure \ref{fig:csidataformat}, we provide an example of UDP packets which carrier a CSI data. Note that, the CSI data is fragmented into 3 packets and it is reassembled by Wireshark. As show in Figure \ref{fig:csidataformat}, the peer address is 0a:19:c6:51:00:12. The MCS of PPDU from which the CSI data is estimated is 9. 

\subsection{PC Software}
\begin{figure}[htbp]
    \centering
    \includegraphics[width=0.45\textwidth]{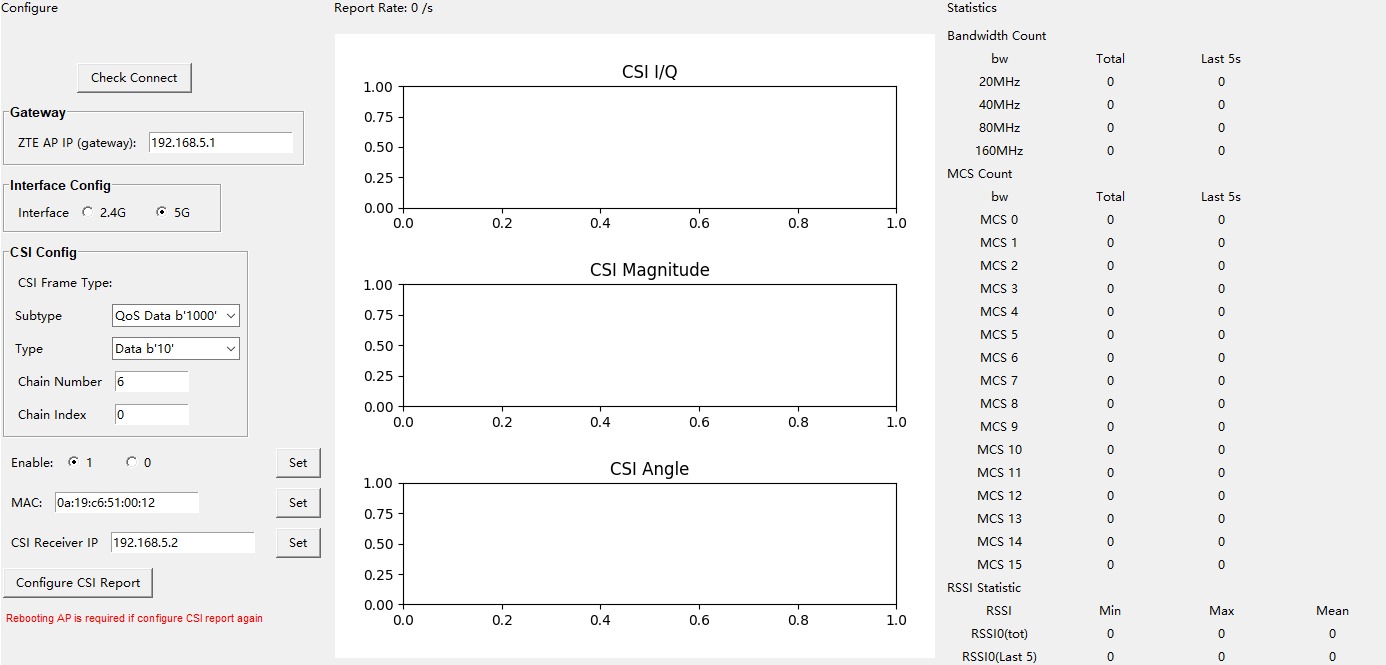}
    \caption{The home page of ZTECSITool.}
    \label{fig:ZTECsiTool}
\end{figure}

We provide then PC software "ZTECSITool" which integrates control, data collection, and statistics. ZTECSITool serves as a best-practice implementation for both the commands and CSI parsing methods presented in this 
\ifdocutohuawei
manual.
\else
paper. 
\fi
Users can adapt and extend its framework to develop their own customized applications.
Figure \ref{fig:ZTECsiTool} shows the main page of the ZTECSITool. The main interface is divided into three sections: CSI Configuration Control, Real-time CSI Visualization, and Statistics Display.
\begin{itemize}
    \item CSI Configuration Control: All UDP commands discussed in Section \ref{sec:config} have been deployed. Just click "Configure CSI Report" button.
    \item Real-time CSI Visualization: The magnitude, phase, and original I/Q data will be drawn. 
    \item Statistics: The number of CSI is counted according to bandwidth and MCS. The average RSSI is also computed. 
\end{itemize}

\subsection{ZTE AP Firmware}

\begin{figure}[htbp]
    \centering
    \includegraphics[width=0.45\textwidth]{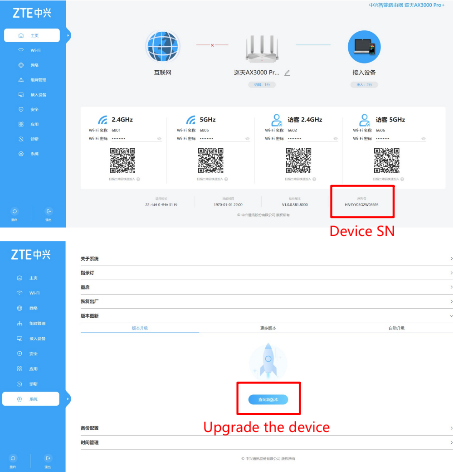}
    \caption{The home page of web pages.}
    \label{fig:webpage}
\end{figure}

Before using ZTECSITool, the firmware of ZTE AX3000 AP should be updated to the CSI-updated experimental version, which supports all functions described in this 
\ifdocutohuawei
manual.
\else
paper. 
\fi
The firmware is deployed on the cloud. User can request this experimental version on the management web page of the ZTE AP. 
To update ZTECSITool firmware, the following step needs to be done:
\begin{enumerate}
    \item Login to the management web page of ZTE AP (by default 192.168.5.1). On the bottom of the main page, record the sequence number.
    \item Send an email to the corresponding author and attach the sequence number. 
    \item Users will receive confirmation from the author and your device will be ready for the experimental test. 
    \item Connect your ZTE AP to Internet through WAN. Login to the management web page. On the system - upgrade, click "request update" bottom. Your ZTE AP will download the experimental firmware and reboot. 
    \item Use ZTECSITool PC software to collection CSI Information.
\end{enumerate}
Figure \ref{fig:webpage} shows the example of web pages. 

\ifdocutohuawei
\else
\section{Conclusion}
In this paper, we present ZTECSITool, a toolkit allowing commercial ZTE AX3000 Series AP uploading CSI information. 
ZTECSITool supports CSI parameters that, to the best of our knowledge, represent the highest known specifications to date, with support for up to 160MHz bandwidth, 512 subcarriers, and 6 chains. We believe that this tool can facilitate the advancement of research in WLAN sensing, especially in areas such as deployment and through-wall detection, and promote the development of the industry
\fi

% or


% \section*{Acknowledgment}


\ifCLASSOPTIONcaptionsoff
  \newpage
\fi



%\bibliographystyle{IEEEtran}
\ifdocutohuawei
\else

\fi

\end{document}